\documentclass[letterpaper, 10 pt, conference]{ieeeconf}  

\IEEEoverridecommandlockouts                              

\usepackage{graphics} 
\usepackage{epsfig} 
\usepackage{mathptmx} 
\usepackage{times} 
\usepackage{amsmath} 
\usepackage{amssymb}  
\usepackage{algorithm}
\usepackage{algpseudocode}

\algrenewcommand\algorithmicrequire{\textbf{Run}}
\algrenewcommand\algorithmicensure{\textbf{Do}}
\algrenewcommand\algorithmicforall{\textbf{for each}}

\algblockdefx{Nest}{EndNest}{}{}
\algtext*{Nest}{}
\algtext*{EndNest}{}

\title{\LARGE \bf
Identification of Black-Box Inverter-Based Resource Control Using Hammerstein-Wiener Models*
}

\author{Aldin Dželo$^{1}$, Amer Mešanović$^{2}$ and Mirsad Cosovic$^{3}$
\thanks{*The authors gratefully acknowledge partial funding by the German Federal Ministry of Education and Research (BMBF) within the Kopernikus Project ENSURE ‘New ENergy grid StructURes for the German Energiewende’.}
\thanks{$^{1}$A. Dželo is with Faculty of Electrical Engineering,
        University of Sarajevo, Bosnia and Herzegovina
        {\tt\small adzelo1@etf.unsa.ba}}%
\thanks{$^{2}$A. Mešanović is with Siemens AG, Munich, Germany
        {\tt\small amer.mesanovic@siemens.com}}%
\thanks{$^{3}$M. Cosovic is with Faculty of Electrical Engineering, University of Sarajevo,
Bosnia and Herzegovina, and the Institute for Artificial Intelligence
Research and Development of Serbia
        {\tt\small mcosovic@etf.unsa.ba}}%
}

\begin{document}

\maketitle
\thispagestyle{empty}
\pagestyle{empty}

\begin{abstract}

The development of more complex inverter-based resources (IBRs) control is becoming essential as a result of the growing share of renewable energy sources in power systems. Given the diverse range of control schemes, grid operators are typically provided with black-box models of IBRs from various equipment manufacturers. As such, they are integrated into simulation models of the entire power system for analysis, and due to their nature, they can only be simulated in the time domain. Other system analysis approaches, like eigenvalue analysis, cannot be applied, making the comprehensive analysis of defined systems more challenging. This work introduces an approach for identification of three-phase IBR models for grid-forming and grid-following inverters using Hammerstein-Wiener models. To this end, we define a simulation framework for the identification process, and select suitable evaluation metrics for the results. Finally, we evaluate the approach on generic grid-forming and grid-following inverter models showing good identification results.

\end{abstract}

\section{INTRODUCTION}

The expansion of inverter-based resources (IBRs) has brought benefits and challenges to grid integration, including the need to ensure sufficient power quality, energy efficiency, and grid resilience \cite{xu2021review}. In order to satisfy newly arisen requirements and create competitive solutions, an increasing number of vendors introduce devices of various architectures to the market. When connecting IBRs to grids, equipment manufacturers are typically required to provide simulation models of their hardware to grid operators. The models are usually provided as black-box models that enable time-domain simulation of the power system. However, it is challenging to use black-box models for other purposes, such as eigenvalue analysis \cite{wang2020scenario}, and they often require different time-steps for simulation or cause numerical instabilities. In contrast, white-box models, implemented in the simulation-software itself and with a known structure, are generally preferred, as they mitigate the issues associated with black-box models. In this work, we introduce an approach for identification of black-box IBR models, based on the Hammerstein-Wiener (HW) model.

Diverse approaches are being used to provide representations of nonlinear IBR dynamics for identification of IBRs, utilizing both time-domain and frequency-domain acquired experimental data \cite{freq_data, 01_xu2017parameter}. Such data is used as a learning foundation for system dynamics in the identification process. The majority of identification approaches use time-domain measured data \cite{02_liu2023research,01_xu2017parameter}. The lack of approaches based on frequency-domain data is attributed to them not being suitable for nonlinear system identification as they are relying on linear techniques \cite{nrel}. Using time-domain data represents a challenge due to the difficulty in interpreting dominant physical features in the data \cite{nrel}. Hence, techniques that combine gathered data and knowledge of the system structure are typically used for IBR model identification. A larger part of studies is based on convergence-based numerical methods that require a priori knowledge of the model's structure (white-box and grey-box model identification). These methods concentrate on accurately identifying previously defined parameters within the structure \cite{03_liu2017two,01_xu2017parameter}. 

Concerning black-box model identification, the most widely used models for representing nonlinear systems are the Nonlinear autoregressive exogenous (NARX) model, HW model and the Artificial neural network (ANN) model. Each option has specific advantages for different setups \cite{qiao2023comparative,zhao2020overview}.
Poor fitness criteria results are often stated as a reason for avoiding HW models in the identification process. While being an important metric for goodness of identification results, its basis lies in the quality of the acquired and proposed validation dataset. Therefore, evaluating the model performance solely on an arbitrarily created dataset is not a good practice \cite{schoukens2009wiener}. 
HW identification has already been undertaken on simplified single-phase structures, but without a general approach that is independent of the operating mode or the type of IBRs \cite{00_patcharaprakiti2010modeling}. It has been shown that there is a possibility of identifying the single-phase inverter system dynamics in steady-state mode \cite{05_patcharaprakiti2012stability}. 

Recent literature on HW IBR identification typically omits the nonlinear components of models in the validation procedure and highlights the lack of a systematic identification process for the general model. Additionaly, the achieved fitness criteria results are inconsistent and only a few of them are considered to be satisfying, even for specific systems~\cite{04_patcharaprakiti2011modeling, 016_abdelsamad2020nonlinear}.


The main novelty of our work lies in introducing, for the first time, a general procedure for HW model-based identification for IBR devices, regardless of their operating mode. 
In contrast to previous works, we consider three-phase models of grid-following (GFL) and grid-forming (GFM) modes of operation, which require a more complex structure for the identified model. A multiple-input multiple-output (MIMO) structure is proposed, consisting of three multiple-input single-output (MISO) subsystems. Furthermore, compared to previous works, the acquired data for the estimation process is processed and transformed from the abc to the dq system, yielding better fitness criteria results for identification accuracy and shortening the identification time. The quality of the developed model is verified by conducting residual analysis, in addition to examining the final prediction error and loss function values.

The paper is structured as follows. In Section II, an overview of GFM and GFL working principles, crucial concerns and requirements are presented in order to propose a general model representation for inverters. The structure and design of the HW model, preparation of the estimation data and the algorithm for the identification procedure are provided in Section III. Numerical results for the algorithm, evaluations for an online-working model and model validation are presented in Section IV. Finally, Section V provides conclusions, highlights encountered challenges, and discusses possible improvements.

\section{GFM AND GFL INVERTER WORKING PRINCIPLES}

There is a multitude of available GFM and GFL control structures, with differences in proposed designs within both industrial and academic communities \cite{gfm_control_strategies}. Since the exact internal model structure of an inverter is not intended to be known in the procedure of black-box identification, it is important to address only the main functions of the inverter. To determine the required functions, key aspects of their general functionalities and grid behavior in GFM and GFL modes are stressed.

In the GFL mode, inverters synchronize with the grid voltages and inject current according to a control logic. The performance of GFL inverters declines in low-strength grids and they cannot operate without an external voltage source \cite{bahrani2024grid}. Gathering simulation data for the identification process requires simulating GFL inverters connected to a voltage source \cite{gfm_control_strategies}. 
When operating in the GFM mode, inverters actively control the voltage magnitude and phase at the connection point. As a result, GFM inverters react almost instantly to system changes, aiding in grid stabilization \cite{bahrani2024grid}. The GFM converter regulates the voltage by directly controlling the power at its output terminals. 

\begin{figure}
    \centering
    \includegraphics[width=\columnwidth]{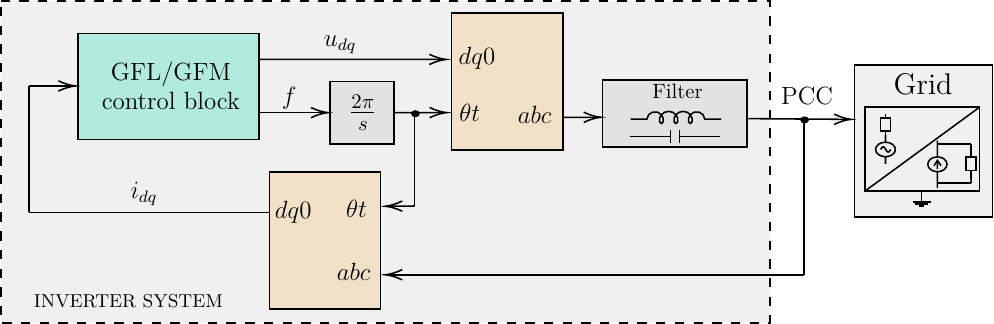}
    \caption{Generalized control structure of GFM/GFL converter.}
    \label{fig:fig1}
\end{figure}

For the utilization of the described structures in our work, a general representation for the inverter in both GFM and GFL modes is considered and depicted in Fig. \ref{fig:fig1}. Based on examples of common inverter model formulations \cite{gfm_control_strategies, du2020modeling}, it has a MIMO structure with terminal currents $i_d$ and $i_q$ in the dq frame as inputs, and terminal voltages $u_d$ and $u_q$ in the dq frame as outputs. Signal $f$ is the controlled frequency of GFM control, or the PLL output frequency of the GFL control. This designed structure enables different classes of control strategies for implementation, such as droop control,  virtual synchronous machine control, virtual oscillator control,  synchronous power control, voltage and current PID-based control \cite{gfm_control_strategies}.     

\section{HAMMERSTEIN-WIENER MODEL DESIGN AND RESULTS}

In HW model design, the black-box system is approximated by the nonlinear HW structure \cite{wills2013identification}. The structure is shown in Fig. \ref{fig:hw_block} and consists of a linear continuous-time system enclosed by an algebraic input and output nonlinearity. The input nonlinear function $f(w, \alpha)$ and the output nonlinear function $h(x, \beta)$ are memoryless, parametrized by vectors $\alpha$ and $\beta$, and may be represented by nonlinearities such as piecewise linear functions or one-dimensional polynomial nonlinearities. Although they are not generally invertible, their first derivatives with respect to the parameter vectors $\alpha$ and $\beta$ exist.

The model identification is based on minimizing the loss function, the weighted sum of the squares of the errors $e(t)$ between the identified model outputs and the black-box measured response. For a model with $n_y$ outputs, the loss function $V(\theta)$ has the following general form \cite{toolbox_ljung}:
\begin{equation}
    V(\theta) = \frac{1}{N} \sum_{t=1}^{N} e^{T}(t,\theta)W(\theta)e(t,\theta)
\end{equation}	
where $N$ is the number of data samples, $e(t,\theta)$ is $n_y \times 1$ error vector at a given time $t$, parameterized by the parameter vector of the HW structure $\theta$, and $W(\theta)$ is the weighting matrix, specified as a positive semidefinite matrix.

Insight into the quality of the identified model is assessed by the fitness quality metrics for accuracy and complexity. The accuracy metric used for the figures presented in this paper is normalized root mean squared error (NRMSE), given as:

\begin{equation}
    \text{NRMSE} = 100 \left( 1-\frac{\left\| y_{\text{measured}} - y_{\text{model}} \right\|}{\left\| y_{\text{measured}}-\overline{y}_{\text{measured}} \right\|} \right)
    \label{eq:NRMSE}
\end{equation}
where $y_{\text{measured}}$ is the measured output data corresponding to $u_{dq}$ and $f$, $\overline{y}_{\text{measured}}$ is mean value for all $y_{\text{measured}}$ channels and $y_{\text{model}}$ is the identified model response. The metric used for complexity comparison is Akaike’s final prediction error (FPE), given as:

\begin{equation}
    \text{FPE} = \text{det}\left( \frac{1}{N}E^{T}E \right)\left( \frac{1+\frac{n_p}{N}}{1-\frac{n_p}{N}} \right)
    \label{eq:FPE}
\end{equation}
where $n_p$ is the number of free parameters in the model, $N$ is the number of samples in the estimation dataset and $E$ is the $N \times n_y$ matrix of prediction errors.

\subsection{Structure of considered models}

We distinguish between two kinds of models: the initial model and the identified model. The initial model, presented in Fig. \ref{fig:initial_model}, is the general external representation of the black-box that is being identified. Establishing the representation of the black-box model is important due to the necessity of knowing the input and output data used for the identification process. Despite the differences in the working principles of inverters in GFM and GFL modes, the same initial model is used for the identification of both types. The justification for this is found within the control structure of both types, where both of them have voltage and frequency as their outputs, either as a regulated values or from the PLL. 

\begin{figure}
    \centering
    \includegraphics[width=\columnwidth]{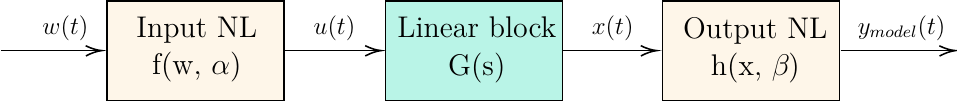}
    \caption{Basic HW structure.}
    \label{fig:hw_block}
\end{figure}

\begin{figure}
    \centering
    \includegraphics[width=\columnwidth]{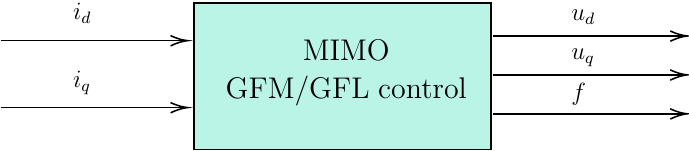}
    \caption{ Assumed initial model MIMO structure for both grid-forming and
grid-following mode.}
    \label{fig:initial_model}
\end{figure}

\begin{figure}
    \centering
    \includegraphics[width=\columnwidth]{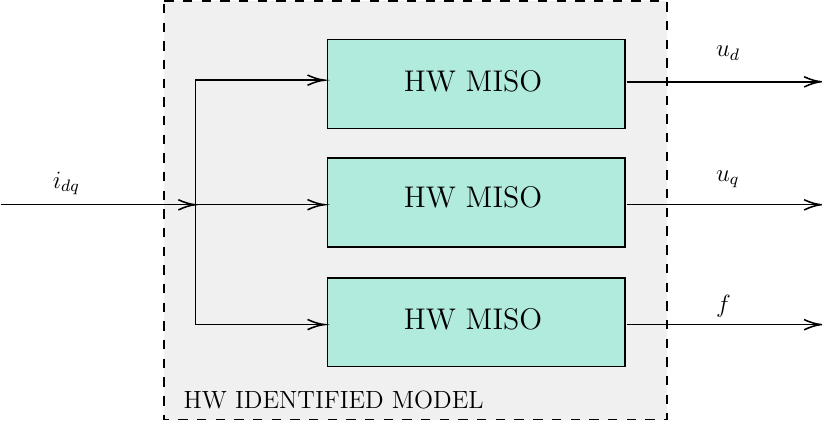}
    \caption{Identified model MIMO structure.}
    \label{fig:fig2}
\end{figure}

The structure of the identified model is shown in Fig. \ref{fig:fig2}, consisting of three basic HW models (see Fig. \ref{fig:hw_block}). Decoupling the initial model's MIMO structure into three MISO structures is justified by greater possibilities regarding design flexibility and reduced computation time for a general solution. The parameters of the MIMO structures are subject to the procedure of fitting with an objective to match the black-box model behavior.

\subsection{Design considerations for the identified model}



\begin{figure}
    \centering
    \includegraphics[width=\columnwidth]{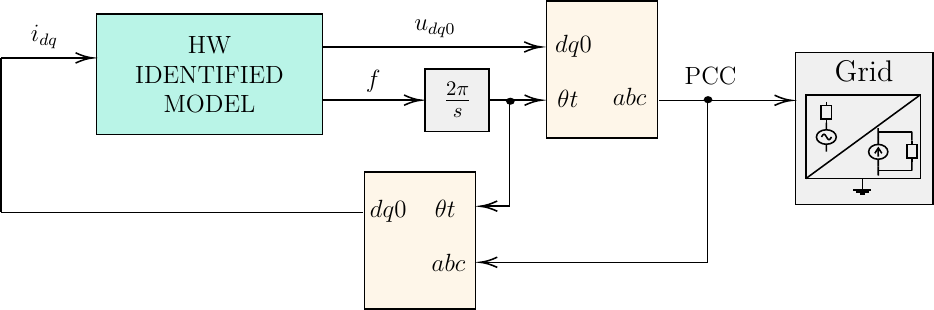}
    \caption{Adaptation of inputs and outputs of the identified system for simulation with other components.}
    \label{fig:fig5}
\end{figure}

As almost all control schemes are in the dq coordinate system, we perform the model identification in this system as well. 
This process first requires the transformation of sinusoidal input signals (voltage and current) into the dq rotating frame, using the measured frequency to calculate the rotating frame angle. Furthermore, when simulating with the identified model from Fig. \ref{fig:fig2}, its inputs and outputs must be transformed into the corresponding frames, as depicted in Fig. \ref{fig:fig5}. 
The described strategy ensures a better fit to the estimation and validation data \cite{toolbox_ljung}, reduces the amount of processed data, and simplifies the modeling and control scheme overall.

Designing the test signals, in light of known inverter applications, requires a well-defined form to excite significant system dynamics. We consider normal operating conditions for the inverter, which include voltage variations between $0.9\,\text{pu}$ and $1.1\,\text{pu}$ and frequency variations within $\pm 0.5\,\text{Hz}$. Expanding to incorporate several operating modes for the identification process leads to complicated solution that will not always provide a complete representation. Therefore, differentiating between several operating modes and having them identified separately enables a faster and simpler approach to find the better solution \cite{isermann2011identification}. The generally favored pseudorandom binary sequence signal cannot be used for this situation due to its unsuitability for detecting input nonlinearities~\cite{isermann2011identification}. 

\subsection{Algorithm for the identification procedure}

Creating a general concept for the identification of GFM and GFL inverters implies having a variety of options for the HW model settings that incorporate possible physical characteristics of the original inverter model. The following settings are considered for variation in this paper:

\begin{itemize}
    \item Degree of nonlinearity estimators for static input/output nonlinearity blocks $f$ and $h$ (see Fig. \ref{fig:hw_block});
    \item Nonlinearity estimators (piecewise linear estimator, one-dimensional polynomial estimator, one-layer sigmoid network estimator and wavelet network estimator);
    \item Numerical search methods used for iterative parameter estimation (Subspace Gauss-Newton least-squares method, Adaptive subspace Gauss-Newton method, Levenberg-Marquardt least squares method, Steepest descent least-squares method);
    \item Order and delay properties of the linear subsystem transfer function.
\end{itemize}

Based on previous observations, the proposed identification procedure is expressed in Algorithm \ref{alg:cap}. After acquiring the necessary estimation and validation data, the identification process consists of iterating through the aforementioned significant parameters. We perform a search by exploring different types of nonlinearity estimators ($nonlinearities$) and iterating over key parameters of the linear block: numerator order ($num$), denumerator order ($denum$), transport delay ($delay$) and nonlinearity degree ($nl\_degree$). The search algorithm is then decided based on the current combination of parameters ($comb$). After the model is obtained with the current combination and search algorithm, it is compared with the current best model in terms of its accuracy and complexity, as defined in \eqref{eq:NRMSE} and \eqref{eq:FPE}. Due to the importance of having a simpler and, therefore, faster system, higher priority is given to systems that are less complex but with relatively lower accuracy. This is accomplished by setting the $\epsilon$ values to favor the complexity criterion. In the end, the best model is validated for quality. If it does not meet all the validation criteria, the second-best model undergoes the same process. This step continues until a model successfully passes the validation.

\begin{algorithm}
\caption{General GFM and GFL identification procedure}\label{alg:cap}
\begin{algorithmic}
\Require the black-box simulation and acquire estimation and validation data
\Require space search:
\ForAll{$nl \in nonlinearities$}
    \ForAll{$comb \in num, denum, delay, nl\_degree$}
                    \State \textbf{Decide} $SearchAlgorithm(est\_data, nl, comb)$
                    \State $model \gets CreateHWModel$
                    \\
                    \State $bestfit \gets bestmodel.fit\_nrmse$
                    \State $bestcomplexity \gets bestmodel.fit\_fpe$
                    \If{$model.fit\_nrmse > bestfit$}
                        \If{$|model.fit\_fpe - bestcomplexity| < \epsilon_{1}$}
                            \State $bestmodel \gets model$
                        \EndIf
                    \ElsIf{$model.fit\_fpe < bestcomplexity$}
                        \If{$|model.fit\_nrmse - bestfit| < \epsilon_{2}$}
                            \State $bestmodel \gets model$
                        \EndIf
                    \EndIf
    \EndFor
\EndFor
\\
\Require model validation procedures from the best model downwards until satisfied (\textit{final prediction error, loss function analysis with validation data and residual analysis})
\end{algorithmic}
\end{algorithm}

\section{NUMERICAL EVALUATIONS}
\begin{figure}
    \centering
    \includegraphics[width=\columnwidth]{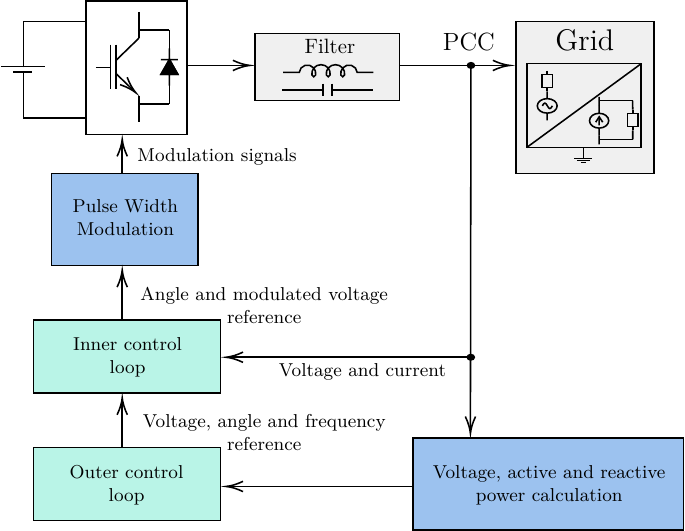}
    \caption{Control structure example of GFM inverter model formulation.}
    \label{fig:gfm_scheme_general}
\end{figure}

\begin{figure}
    \centering
    \includegraphics[width=\columnwidth]{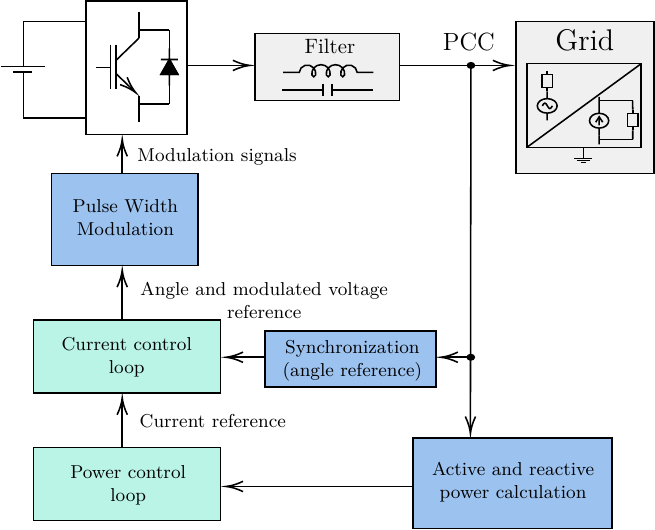}
    \caption{Control structure example of GFL inverter model formulation.}
    \label{fig:gfl_scheme_general}
\end{figure}

To evaluate the applicability of the described algorithm, we consider examples of common GFM and GFL inverter model formulations \cite{gfm_control_strategies, bahrani2024grid, du2020modeling}, depicted in Fig. \ref{fig:gfm_scheme_general} and Fig. \ref{fig:gfl_scheme_general}. By treating them as black-box models, these inverters are simulated with only their input and output time-domain data being acquired. The nature of the inputs and outputs is the same as in Fig. \ref{fig:initial_model}, and each output is subjected to the space search within its corresponding HW MISO system, as depicted in Fig. \ref{fig:fig2}. 

\begin{figure}
    \centering
    \includegraphics[width=\columnwidth]{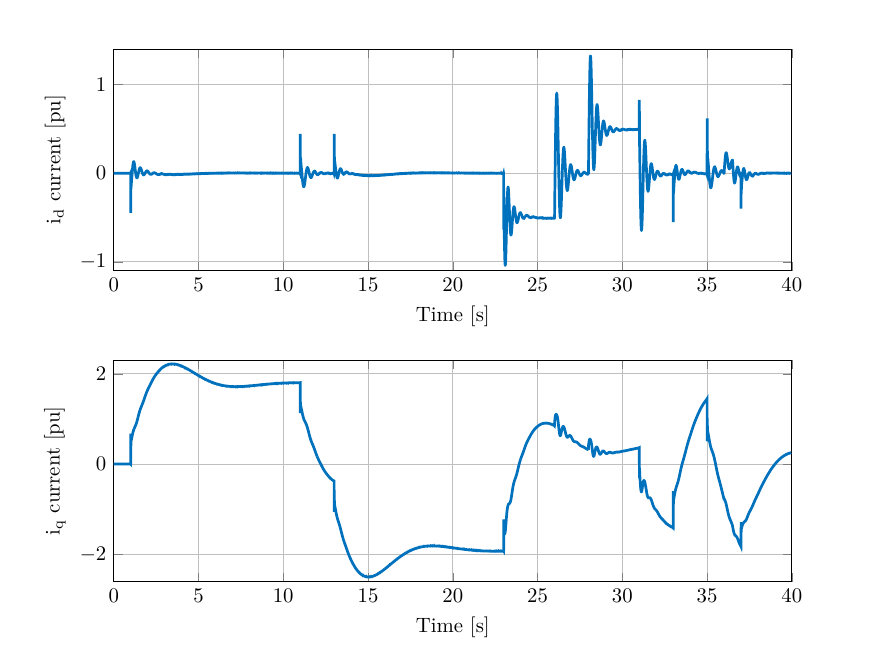}
    \caption{Inputs (GFM mode) estimation data for normal operating mode.}
    \label{fig:est_input}
\end{figure}

\begin{figure}
    \centering
    \includegraphics[width=\columnwidth]{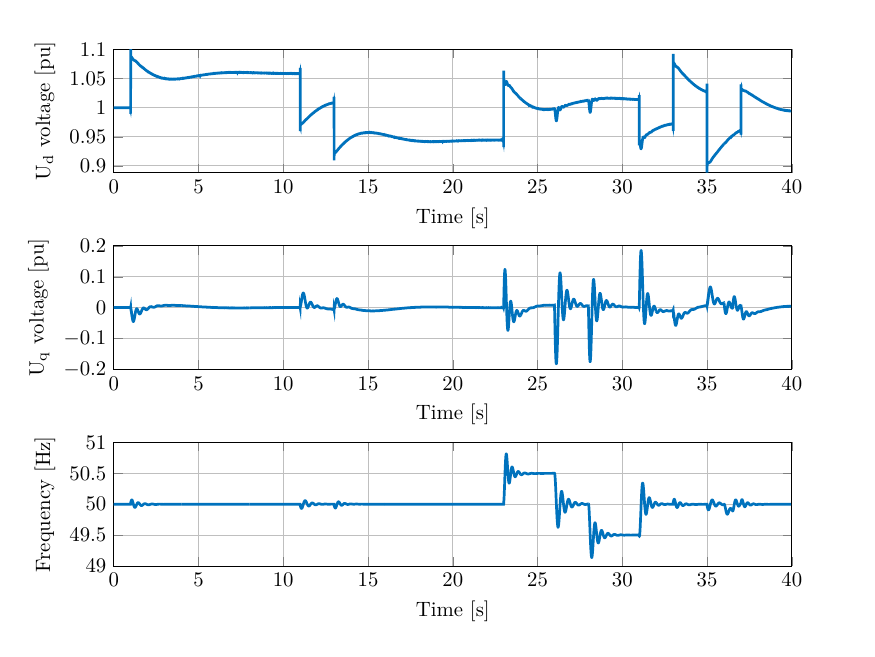}
    \caption{Outputs (GFM mode) estimation data for normal operating mode.}
    \label{fig:est_output}
\end{figure}

The proposed test signals are created by simulating the scheme depicted in Fig. \ref{fig:fig1} and are acquired at the point of common coupling (PCC). The test signals for the estimation data of the normal operating mode for grid-forming inverters are shown in Fig. \ref{fig:est_input} and Fig. \ref{fig:est_output}. Both voltage and frequency variations are present, allowing the system to achieve a steady state and satisfy time limitations due to the computation time criteria. 

The identification results include the number of possible variations of the identified HW model for the examined system, with an NRMSE fit greater than 92\,\%. A fit threshold of 92\,\% was arbitrarily chosen, as it satisfactorily captures the dynamics, as confirmed through the validation procedure. Although the best model solution has the optimal accuracy-to-complexity ratio based on the established metrics, the number of other possible model variations are presented to provide a better understanding of the proposed algorithm's effectiveness. Additionally, this offers insight into alternative options if different accuracy-to-complexity ratio preferences are prioritized.

\subsection{Hammerstein-Wiener GFM model}\label{AA}




During an extensive numerical space search, the following results were attained considering estimation data:
\begin{itemize}
    \item 760 possible HW model structures for frequency MISO block solution with fits greater than 92\,\%.
    \item 187 possible HW model structures for $U_d$ voltage MISO block solution with fits greater than 92\,\%.
    \item 61 possible HW model structures for $U_q$ voltage MISO block solution with fits greater than 92\,\%.
\end{itemize}

\begin{figure}
    \centering
    \includegraphics[width=\columnwidth]{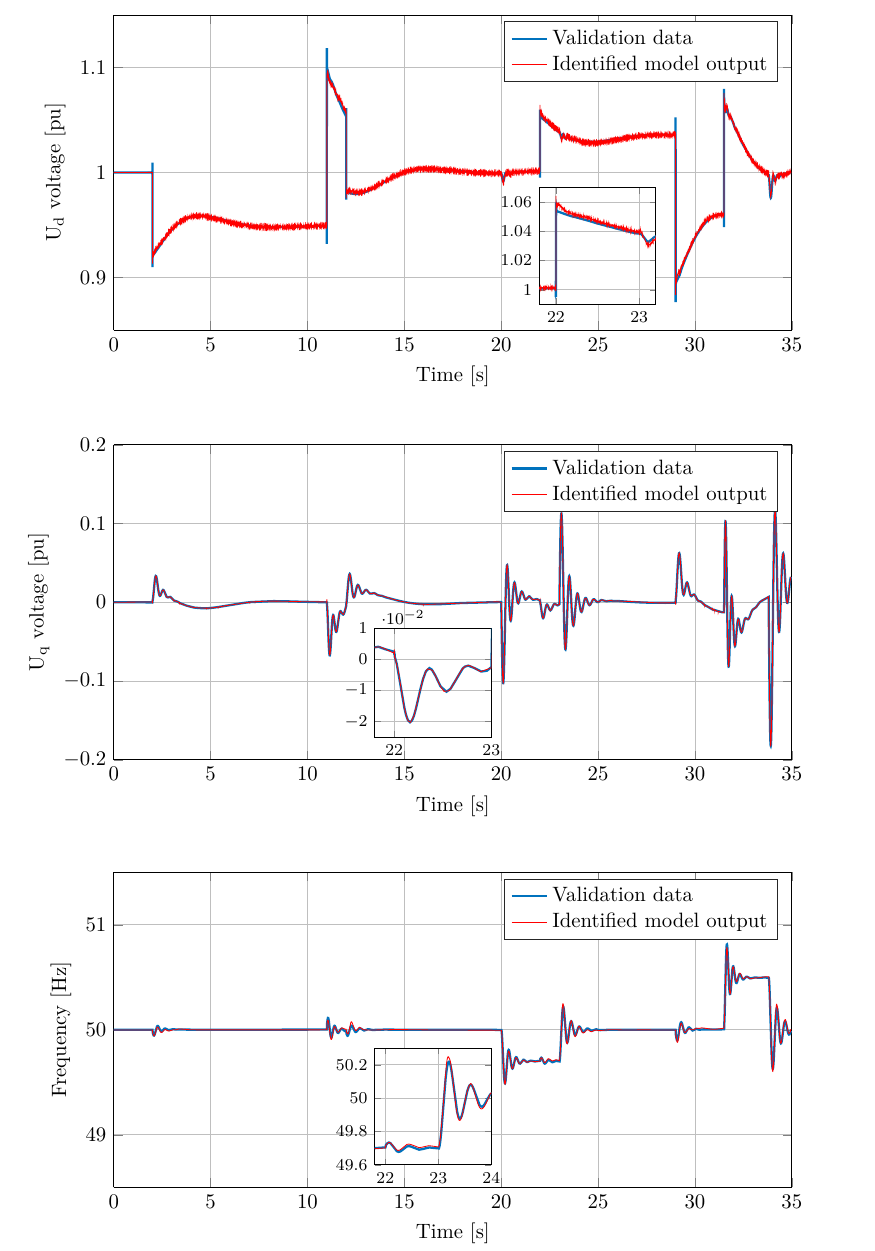}
    \caption{Comparison between actual model and identified model outputs in grid-forming mode.}
    \label{fig:valtest}
\end{figure}

The comparison between actual model and identified model outputs for the validation data set is shown in Fig. \ref{fig:valtest}. The best solution obtained has the following fits for the estimation data: $97.85\,\%$ (frequency), $98.05\,\%$ ($U_d$ voltage) and $99.53\,\%$ ($U_q$ voltage). As for the observed validation data fits, the results are as follows: $95.23\,\%$ (frequency), $96.89\,\%$ ($U_d$ voltage) and $99.47\,\%$ ($U_q$ voltage).

\subsection{Hammerstein-Wiener GFL model}

During an extensive numerical space search, the following results were attained considering estimation data:
\begin{itemize}
    \item 473 possible HW model structures for frequency MISO block solution with fits greater than 92\,\%.
    \item 49 possible HW model structures for $U_d$ voltage MISO block solution with fits greater than 92\,\%.
    \item 56 possible HW model structures for $U_q$ voltage MISO block solution with fits greater than 92\,\%.
\end{itemize}

\begin{figure}
    \centering
    \includegraphics[width=\columnwidth]{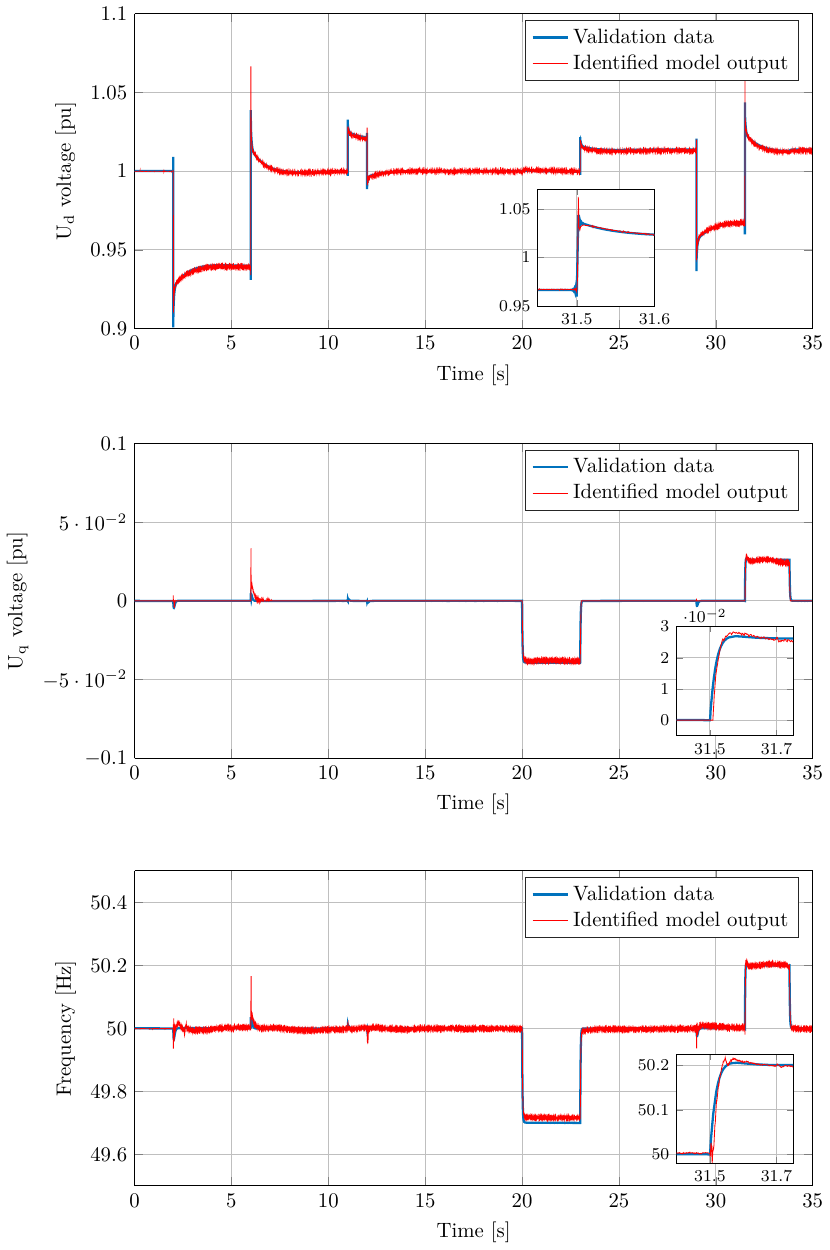}
    \caption{Comparison between actual model and identified model outputs in grid-following mode.}
    \label{fig:valtest_gfl}
\end{figure}

The comparison between actual model and identified model outputs for the validation data set is shown in Fig. \ref{fig:valtest_gfl}. The best solution obtained has the following fits for the estimation data: $96.76\,\%$ (frequency), $97.36\,\%$ ($U_d$ voltage) and $98.31\,\%$ ($U_q$ voltage). As for the observed validation data fits, the results are as follows: $92.79\,\%$ (frequency), $95.84\,\%$ ($U_d$ voltage) and $93.74\,\%$ ($U_q$ voltage).

\subsection{Online-working model design}

Previously derived models should not only be accurate for offline experiments, where both the estimation and validation data is acquired before the identification process. Instead, they become significant if they are workable in ongoing grid simulations. 

\begin{figure}
    \centering
    \includegraphics[width=\columnwidth]{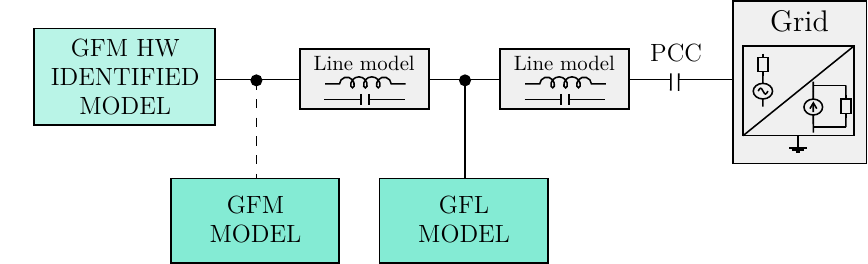}
    \caption{Validation microgrid scheme for model in simulation.}
    \label{fig:microgrid}
\end{figure}

\begin{figure}
    \centering
    \includegraphics[width=\columnwidth]{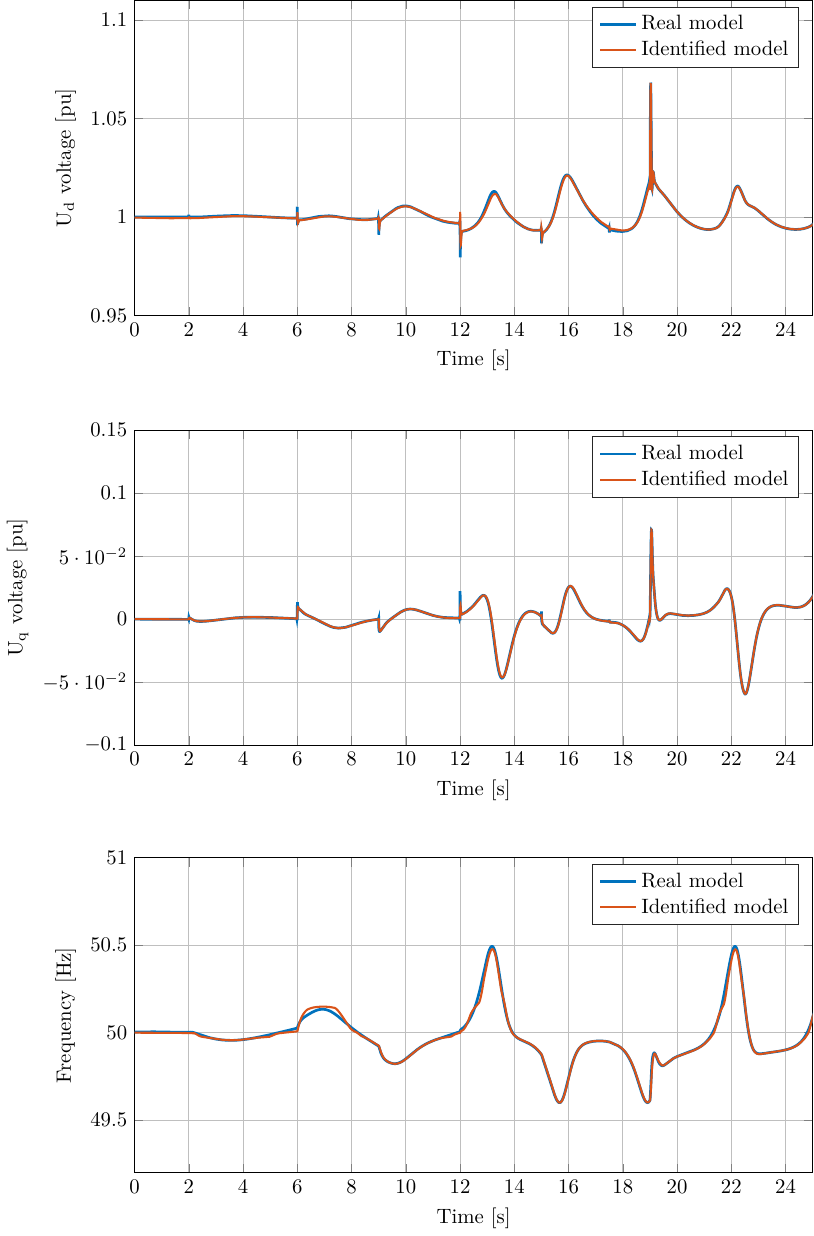}
    \caption{Simulation outputs comparison between actual and adapted identified model.}
    \label{fig:sim_solution_fig}
\end{figure}

A simulation microgrid experiment was conducted with varied specified voltage at the PCC (see Fig. \ref{fig:microgrid}) and the results are presented in Fig. \ref{fig:sim_solution_fig}. 
It can be concluded that the identification model achieves satisfactory results in terms of stability and demonstrates good accuracy.

\subsection{Residual analysis}

To gain confidence in the identified model, one can ensure that model errors are distinguished from disturbances through residual analysis, which includes autocorrelation and cross-correlation tests \cite{ljung1995system}. Autocorrelation and cross-correlation tests provide insights into the model fidelity regarding the candidate model structure and account for possible disturbances (disturbance path and input-output path respectively). 





High autocorrelation test results have been observed for the HW identification \cite{jespersen2023hammerstein}. This is because of the high disturbance level at the model connection points with the rest of system. 

\begin{figure}
    \centering
    \includegraphics[width=\columnwidth]{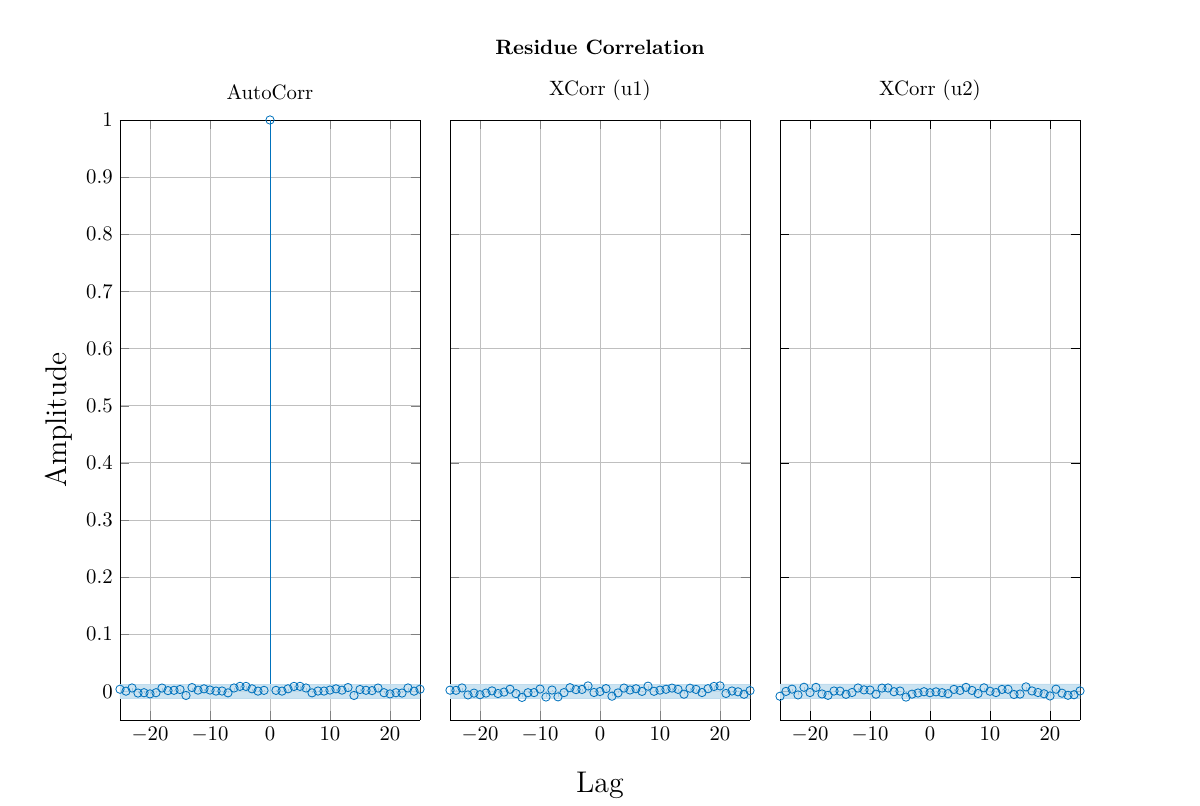}
    \caption{Residual analysis results for the identified model.}
    \label{fig:good}
\end{figure}

By addressing the problem of disturbance path and incorporating filter dynamics, the residual analysis for the identified model gets as in Fig. \ref{fig:good}. The confidence interval, which corresponds to the range of statistically insignificant residual values for the system, is marked by blue area and it is specified with a probability of 99\,\%. The proposed input-output path in the model structure is acceptable because the cross-correlation residuals values lie within the confidence boundaries. The autocorrelation function of the residuals is within the confidence area except at lag 0, where the signal is correlated with itself. It can be concluded that the identified model achieves satisfactory results.



\section{CONCLUSIONS}

In this work, we propose an algorithm for automatic model identification of black-box IBR models based on HW model identification. Key aspects of the identification process are determined by analyzing GFM and GFL inverter functionalities and grid behavior. We adopt a MIMO structure representing the inverter system, independent of its mode. By selecting a loss function, and accuracy and complexity metrics, we lay the foundation for developing a simpler and faster substitution model. Several decisions were made to achieve a more general and improved result: (i) three-phase architectures for both GFM and GFL are considered; (ii) the original MIMO model structure is decoupled into MISO subsystems for each output; (iii) identification data is transformed into the dq system; (iv) residual analysis is introduced as part of the model validation process. The proposed identification algorithm integrates the paper's considerations and is verified with GFM and GFL model examples. A microgrid with GFM and GFL inverters further confirmed the model's stability and accuracy.

We conclude that the proposed solution is independent of subsystem variations in inverter implementations. However, it cannot reliably link the physical characteristics of linear or nonlinear blocks with actual inverter subsystems. Nonetheless, the overall model expands analysis options, enabling studies such as singular value analysis, advanced stability analysis, and frequency domain analysis, which were previously unavailable. As noted in \cite{bahrani2024grid}, a key issue in inverter network stability studies is the lack of complete models and standardization. Traditional methods struggle to assess small-signal stability in networks with mixed IBRs and synchronous machines. We believe this identification framework will deliver an upturn in this matter.

For future work, the challenge of interpreting the physical characteristics of the identified blocks could be addressed. Additionally, the question of the complexity metric remains unresolved, as it is uncertain whether different system simplification techniques will yield better general results.

\addtolength{\textheight}{-12cm}   











\bibliographystyle{IEEEtran}
\bibliography{literatura}

\end{document}